\documentclass[a4paper,twocolumn,11pt,accepted=2017-05-09,dvipsnames]{quantumarticle}
\pdfoutput=1
\usepackage[utf8]{inputenc}
\usepackage[english]{babel}
\usepackage[T1]{fontenc}
\usepackage{amsmath}
\usepackage{hyperref}
\usepackage[numbers]{natbib}

\usepackage{tikz}
\usepackage{lipsum}
\usepackage{amssymb}
\usepackage[colorinlistoftodos, color=green!40, prependcaption]{todonotes}
\usepackage{amsthm}
\usepackage{mathtools}
\usepackage{physics}
\usepackage{diagbox}
\usepackage{makecell}
\usepackage[left=20mm,right=20mm,top=30mm, columnsep=15pt]{geometry}
\usepackage{adjustbox}
\usepackage{relsize}
\usepackage{placeins}
\usepackage{hyperref}
\usepackage{extarrows}
\usepackage{csquotes}
\usepackage[]{graphicx}
\usepackage{float}
\usepackage{qcircuit}
\usepackage{pgfplots}
\usepackage{amsmath}
\usepackage{braket}
\usepackage{multirow}
\usepackage{verbatim}
\usepackage{dsfont}
\usepackage{bm}
\usepackage[shortlabels]{enumitem}
\usepackage{rotating}
\numberwithin{equation}{section}

\definecolor{USI}{RGB}{254,226,124}
\definecolor{Wien}{RGB}{0,99,166}
\definecolor{QSIT}{RGB}{0,152,155}
\definecolor{qcol}{RGB}{82,41,125}
\definecolor{qsitcol}{HTML}{00989a}
\definecolor{Bblue}{rgb}{0.19, 0.55, 0.91}

\newcommand{\proj}[1]{|#1\rangle\langle#1|}
\begin{document}
\title{Wigner’s friend’s memory and the no-signaling principle}

\author{Veronika Baumann,}
    %\email[Correspondence email address: ]{martin.renner@univie.ac.at}% Your name
    \email{veronika.baumann@oeaw.ac.at}% Your name
    \affiliation{Institute for Quantum Optics and Quantum Information (IQOQI), Austrian Academy of Sciences, Boltzmanngasse 3, 1090 Vienna, Austria}
    \affiliation{Atominstitut, Technische Universität Wien, 1020 Vienna, Austria}
    
\author{\v{C}aslav Brukner}
    %\email[Correspondence email address: ]{martin.renner@univie.ac.at}% Your name
    \email{caslav.brukner@univie.ac.at}% Your name
    \affiliation{Institute for Quantum Optics and Quantum Information (IQOQI), Austrian Academy of Sciences, Boltzmanngasse 3, 1090 Vienna, Austria}
     \affiliation{University of Vienna, Faculty of Physics, Boltzmanngasse 5, 1090 Vienna, Austria}

%\date{\today}

\begin{abstract}

The Wigner's friend experiment is a thought experiment in which a so-called superobserver (Wigner) observes another observer (the friend) who has performed a quantum measurement on a physical system. In this setup Wigner treats the friend, the system and potentially other degrees of freedom involved in the friend's measurement as one joint quantum system.
In general, Wigner's measurement changes the internal record of the friend's measurement result such that after the measurement by the superobserver the result stored in the observer's memory register is no longer the same as the result the friend obtained initially, i.e. before she was measured by Wigner. Here, we show that any awareness by the friend of this change of her memory, which can be modeled by an additional register storing the information about the change, conflicts with the no-signaling condition in extended Wigner-friend scenarios. \\
\end{abstract}

\maketitle

\section{Introduction}
\label{introduction}

In the famous Wigner's friend thought experiment~\cite{wigner1963problem} one considers the observation of an observer in order to illustrate the quantum measurement problem~ \cite{maudlin1995three,bub2010two,brukner2017quantum}. It comprises an observer -- called Wigner's friend (F) --, who measures a quantum system (S), as well as a so-called superobserver -- Wigner (W) -- who performs a measurement on the joint quantum system of S and F. It is assumed that Wigner's friend uses Lüder's rule after obtaining a definite outcome for her measurement. However, provided that the joint system S+F is sufficiently isolated, Wigner describes the friend's measurement via unitary dynamics and therefore assigns an entangled state to the composite system. This difference in description of the same process, namely the friend's interaction with her measured system, is called the Wigner's friend paradox. It implies that Wigner and his friend would make disagreeing predictions about subsequent measurements on the joint system S+F, provided that both observers are assumed to be able to reason about the setup they are part of. 

In recent years new scenarios combining Wigner's-friend-type experiments with various non-locality setups~\cite{Zukowski2020,Leegwater:2018aa,brukner2018no,frauchiger2018quantum,brukner2017quantum} have been proposed. These extended Wigner's friend experiments have been used to devise new powerful no-go theorems for a classical notion of objectivity in quantum theory.
More concretely, a set of assumptions called local friendliness~\cite{Bong2020}  --  namely, that the superobservers’ and observers’ results are both ``objective facts'', ``locality'', ``freedom of choice'', and ``universality of quantum theory'' -- cannot all hold in extended Wigner's friend setups. The last assumption asserts that the unitary description of the friend's measurement correctly predicts Winger's observed statistics. The performed proof-of-principle tests of the no-go theorems confirm the quantum mechanical correlations for the results of the superobservers. However, the ``friend'' was in that case modeled by  a single photon~\cite{proietti2019experimental,Bong2020}, for which the assignment of the notion ``observer'' or ``observation'' has little physical meaning; compare Ref.~\cite{brukner2021qubits}.  
{The significance of the new no-go theorems lies precisely in the point that they question the objectivity of observable facts which observers perceive, be aware of or may have knowledge about, and not the counterfactual properties of simple quantum systems, which are the subject of Bell's theorem. Moreover, by the natural assumption that the friend has a definite perception of her observational facts, just as Wigner has of his, we are put in the position of considering the friend's observations as "hidden variables" from Wigner's point of view, not described by quantum theory. This is different from Bell's introduction of hidden variables, which refer to unobservable or counterfactual properties of simple systems. Attributing notions like ``perception'' and ``awareness'' to the friend
was also the motivation for } an extended Wigner's friend experiment involving a human level AI on a quantum computer taking the role of the friend~\cite{wiseman2023thoughtful}. There the friend is assumed to have ``internal thoughts'', which in~\cite{wiseman2023thoughtful} is assumed to constitute a sufficient condition to be an observer. In this work we address the question: What properties must the internal thoughts of observers who are described by a superposition of perception states by other observers have in order to be compatible with known physical principles?

\begin{figure}[hbt!]
\begin{tikzpicture}[scale=0.58]

  \draw[thick,] (-4,-3) rectangle (5.5,3);
  \node[fill=YellowGreen!30, draw=YellowGreen, thick] (s) at (-3,0) {$S$};

   \draw[->,>=latex, thick,] (s) -- node [midway,above] {$\ket{\phi}_S$} (-0.5,0);
      
  \node[fill=PineGreen!20, draw=PineGreen, thick, rounded corners=3pt] (r1) at (3.75,1.25) {$\ket{0}_S\ket{\bm 0}_F\,$};
  \node[fill=PineGreen!20, draw=PineGreen, thick, rounded corners=3pt] (r2) at (3.75,-1.25) {$\ket{1}_S\ket{\bm 1}_F\,$};
  \draw[->,>=latex,thick, draw=gray] (1,0) to[bend left] (r1);
  \draw[->,>=latex,thick, draw=gray] (1,0) to[bend right] (r2);
  
  \node[] (m) at (1,0) {\includegraphics[scale=0.15]{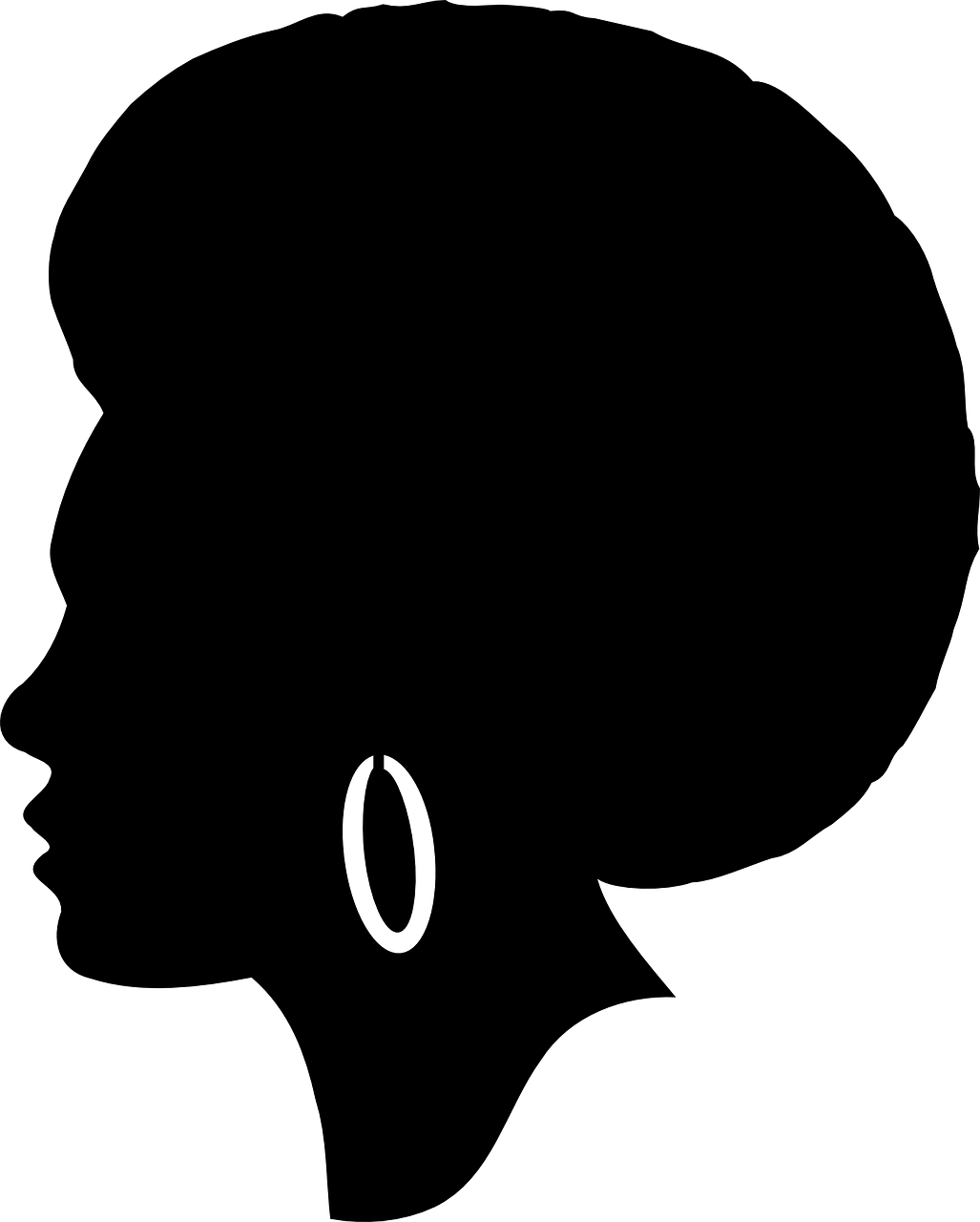}};
  \node[] (F) at (1,-2) {F};
  %\node[fill=Bblue!20, draw=Wien, thick, rounded corners=3pt] (x1) at (12.5,2) {$\ket{W=1}_{SF}$};
  %\node[fill=Bblue!20, draw=Wien, thick, rounded corners=3pt] (x2) at (12.5,-2) {$\ket{W=2}_{SF}$};
   %\draw[->,>=latex,shorten >=2pt,shorten <=2pt, thick, draw=gray] (9,0) to[bend left] (x1);
  %\draw[->,>=latex,shorten >=2pt,shorten <=2pt, thick, draw=gray] (9,0) to[bend right] (x2);

  \draw[->,>=latex,thick] (5.5,0) --node [midway,above]{$\ket{\Phi}_{SF}$} (8,0);

  \node[] (M) at (9,0) {\includegraphics[scale=0.05]{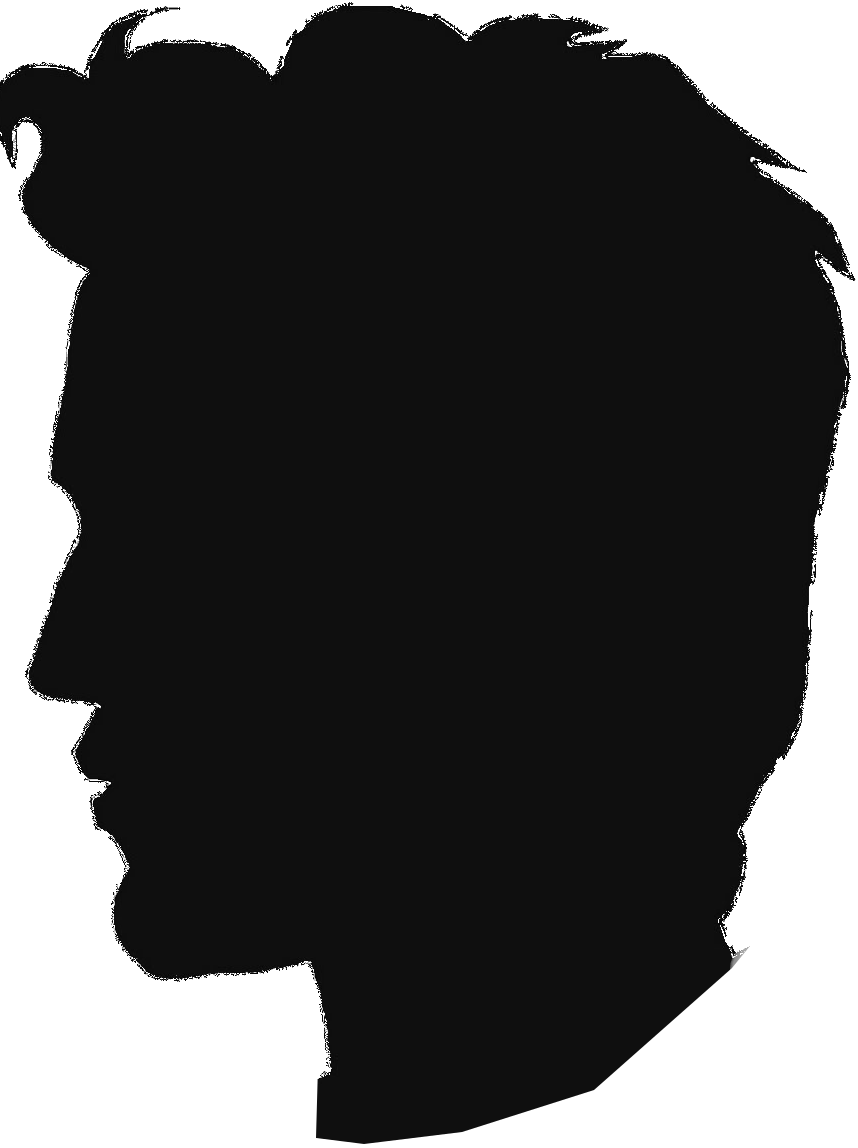}};
    \node[] (W) at (9,-2) {W};
 \end{tikzpicture}
 \caption{In the original Wigner's friend setup the source emits a quantum system in state $\ket{\phi}_S$, which is measured by the friend in the computational basis at time $t_F$. The result observed by the friend in stored in some memory register $\ket{\cdot}_F$. Then Wigner performs his measurement on both the system and the friend's memory register at time $t_W$. His measurement will in general alter the friend's memory register meaning that what is stored in her memory at time $t_2>t_W$ is different from what was encoded in her memory at time $t_1<t_W$.
 }
    \label{Simple Wigner}
\end{figure}
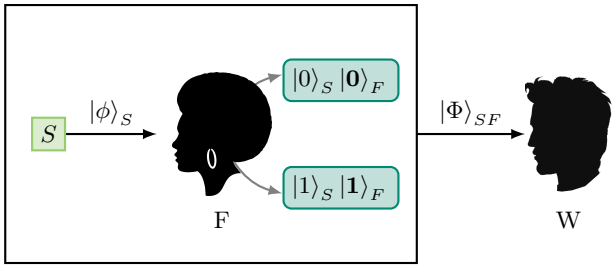

Wigner's measurement will, in general, alter the friend's perception of her measurement result, i.e. state of the friend's memory which encodes the result~\cite{deutsch1985quantum,baumann2018formalisms,cavalcanti2021view,baumann2020wigner}. The paper~\cite{allard2020no} considers the perception state of the friend in the original Wigner's friend experiment before and after Wigner's measurement, see Fig.~\ref{Simple Wigner}. It proves that no joint probability distribution of the friend's perceptions before and after Wigner's measurement can agree with the unitary description endorsed by Wigner and be a convex linear function of the initial state. 
This linearity assumption was criticized as too strong in Ref.~\cite{schmid2023review}. In this work we drop this assumption, which removes a major limitation on a joint probability of the friend's perceived result before and after Wigner's measurement. This, in turn, allows for more general considerations about the conditional probabilities of friend's perception at two different times, i.e. the changes of her memory due to Wigner's measurement. \\

Here, we prove the incompatibility between general models of the observer's memory and the no-signaling condition. More concretely, we show that if the friend can be aware of the change in her memory  in an extended Wigner-friend scenario (i.e. if she has an additional memory register that can record information about changes in the memory registers storing the measurement result), this can lead to signaling. A distant observer --Bob (B)-- could signal to the friend by choosing different measurement settings in his lab. We then conclude that compatibility with the no-signaling condition requires that observers \emph{cannot} be aware of the change in their internal memory induced by a superobserver.  

The paper is structured as follows. In section~\ref{Changes} we reexamine the simple Wigner's friend experiment and consider what the changes of the friend's perception could look like. In section~\ref{ConfSig} we then apply this to an extended Wigner's friend setup and give an example of a protocol that would be signaling, if the friend was aware of the change of her memory. We present our conclusions in section~\ref{Conclusions}.

\section{Changes of the friend's memory}
\label{Changes}

In the following, we consider probabilities for the results $f$ stored in the friend's memory at a certain time $t$.
{The computational basis states of this memory registers are assumed to correspond to the friend perceiving the respective result, see Ref.~\cite{brukner2021qubits}. This result is, however, \emph{not} accessible to Wigner unless he performs the corresponding measurement on his friend. We can nevertheless reason about the probabilities of the friend's perception, if Wigner does not perform such a measurement or even if he undoes the measurement interaction and returns his friend and the system to their pre-measurement state~\cite{gao2019quantum}.} Here, we describe the situation where Wigner makes a joint measurement on the system and the friend, and compare the probabilities for the results $f$ \emph{encoded in the friend's memory} before and after Wigner's measurement. To this end, we further assume that unitary quantum theory prescribes the assignments of a probability for observations at each individual point in time and that an observer, who applies this prescription, will predict probabilities that are consistent with the relative frequencies in an actual experiment. {This means that the probability of an observer $O$ perceiving outcome $x$ at time $t_j$ is given by $p(x_j)= \bra{\Psi(t_j)}  \mathds{1}\otimes\proj{X_x}_O \ket{\Psi(t_j)} $ where $\ket{\Psi(t_j)}$ is the unitarily evolved total state involving the measured system as well as the observer. The states $\ket{X_x}$ correspond to the state of observer $O$ perceiving outcome $x$.  They are the computational basis states of the respective memory space $\mathcal{H_O}$. A superobserver, Wigner, \emph{can} measure the observer, the friend, in said basis, which is usually referred to as ``asking the friend which result she observed'', but if he chooses not to do so, the result stored in $F$'s memory is empirically inaccessible to Wigner due to the assumed isolation of the joint system S+F.}
If the friend measures a qubit in the computational basis at time $t_F$ and is then measured by Wigner at time $t_W$, the probabilities for the results $\bm 0$ and $\bm 1$ being stored in her memory are, in general, different at a time $t_1<t_W$ and $t_2>t_W$. 

Suppose that at time $t_W$ Wigner performs a measurement given by states $\ket{W=1}= a\ket{0}_S\ket{0}_F+b\ket{1}_S\ket{1}_F$, $\ket{W=2}=b^*\ket{0}_S\ket{0}_F-a^*\ket{1}_S\ket{1}_F$ and their orthogonal complement $\Pi_{\perp}=\mathds{1}-\proj{W=1}-\proj{W=2}$. For an initial system state $\ket{\phi}_S=\alpha \ket{0} + \beta \ket{1}$ we obtain
\begin{align}
p(f_1=0)=& |\alpha|^2  \\
p(f_1=1)=& |\beta|^2   \\
\label{pf20} 
p(f_2=0)=& |\alpha|^2 (|a|^4+|b|^4)+2 |\beta|^2|a|^2|b|^2 \\ \nonumber
	 &+ 2|\alpha| |\beta| (|a|^3|b| - |a||b|^3)\cos{\theta} \\ 
\label{pf21}
p(f_2=1)=& |\beta|^2 (|a|^4+|b|^4)+2 |\alpha|^2|a|^2|b|^2 \\ \nonumber
	&- 2|\alpha| |\beta| (|a|^3|b| - |a||b|^3)\cos{\theta},
\end{align}
where $\theta$ is a relative phase factor, see Appendix~\ref{app:Simple} for details. Here, we denote by $p(f_1=i)$ the probability of result $i$ being stored in the friend's memory at time $t_1$ and by $p(f_2=i)$ the probability that at time $t_2$ the result $i$ is stored in the friend's memory. Both these probabilities individually can be empirically accessible to Wigner if in multiple runs of the experiment he asks the friend which result is stored in her memory at the corresponding instant of time. Note that he cannot access both probabilities in the same runs of the experiment, since $p(f_2)$ refers to the result stored in the friend's memory after Wigner performed the measurement given by $\proj{W=1}, \proj{W=2}$ and $\Pi_{\perp}$ but did \emph{not} measure which result the friend observed at $t_1$. Wigner measuring $p(f_1)$ corresponds to a different scenario including a measurement in the perception basis \{$\ket{\bm{f}}_F$\}, which will, in general, alter  subsequently observed statistics.

\subsection{Memory changes in the simplest Wigner's friend setup}
\label{Changes_S}

\begin{figure}[hbt!]
\begin{tikzpicture}[scale=0.75]

  \node[] (a) at (-4,1) {(a)};
  \node[] (0f1) at (-3,1) {\color{PineGreen}$\bm 0$};
  \node[] (1f1) at (-3,-1) {\color{PineGreen}$\bm 1$};
  \node[] (0f2) at (-1,1) {\color{qcol}$\bm 0$};
  \node[] (1f2) at (-1,-1) {\color{qcol}$\bm 1$};
  
   \draw[->,>=latex, thick,] (1f1) -- (0f2);
   \draw[->,>=latex, thick,] (0f1) --  (1f2);
   \node[fill=white] (q)at (-2,0){$q$};
   \draw[->,>=latex, thick,] (0f1) -- node [midway,above] {$1-q$} (0f2);
  \draw[->,>=latex, thick,] (1f1) -- node [midway,below] {$1-q$} (1f2);
  
   \node[] (b) at (2,1) {(b)};
  \node[] (0f2m) at (5,1) {\color{qcol}$\bm 0$};
  \node[] (1f2m) at (5,-1) {\color{qcol}$\bm 1$};
  \node[] (0f1m) at (3,1) {\color{PineGreen}$\bm 0$};
  \node[] (1f1m) at (3,-1) {\color{PineGreen}$\bm 1$};
  
   \draw[->,>=latex, thick,] (1f1m) -- (0f2m);
   \draw[->,>=latex, thick,] (0f1m) --  (1f2m);
   \node[] (q)at (3.9,0.6){$q^0$};
   \node[] (q)at (3.9,-0.5){$q^1$};
   \draw[->,>=latex, thick,] (0f1m) -- node [midway,above] {$1-q^0$} (0f2m);
  \draw[->,>=latex, thick,] (1f1m) -- node [midway,below] {$1-q^1$} (1f2m);
  
 \end{tikzpicture}
 \caption{The results stored in the friend's memory at time $t_1$ are shown in \textcolor{PineGreen}{green}, while the memory entries at time $t_2$ are depicted in \textcolor{qcol}{purple}. (a): Both results $\bm 0$ and $\bm 1$ are flipped with probability $q$ and stay the same with probability $1-q$. (b): With probability $q^0$ an initially observed result $\bm 0$ gets flipped to result $\bm 1$. With probability $1-q^0$ the stored result $\bm 0$ is not changed by Wigner's measurement. Similarly an initially stored $\bm1$ gets altered with probability $q^1$ and remains unchanged with probability $1-q^1$.}
    \label{Flip}
\end{figure}
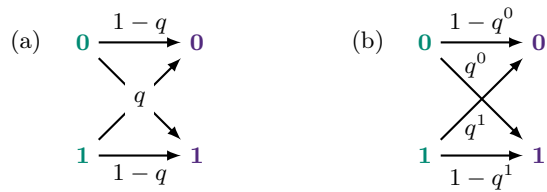

As implied by Ref.~\cite{allard2020no} we cannot model the change of the friend's memory by parameters that are independent of the initial state. The memory change can be mathematically described by
\begin{align}
\label{p2_0}
p(f_2)= \sum_{f_1}p(f_1)p(f_2|f_1).
\end{align}
Let's assume $p(f_2|f_1)=q$ if $f_2\neq f_1$ corresponds to the probability of the result stored in the friend's memory being flipped during Wigner's measurement, see Fig.~\ref{Flip}(a). More concretely, using this single parameter $q$ (the flipping probability) we can write $p(f_2)$ for the setup in Fig.~\ref{Simple Wigner} as follows:
\begin{align}
\label{p2_0}
p(f_2=0)= |\alpha |^2 (1-q)+|\beta |^2q \\
\label{p2_1}
p(f_2=1)=|\alpha |^2 q+|\beta |^2 (1-q).
\end{align}
Since these probabilities also have to satisfy Eqs.~\eqref{pf20} and~\eqref{pf21} we obtain the following system of equations
\begin{align}
\label{sys1_1}
|\alpha |^2 &-q(|\alpha |^2-|\beta |^2)=\\ 
& \qquad |\alpha|^2 (|a|^4+|b|^4)+2 |\beta|^2|a|^2|b|^2 +2\chi \nonumber \\
\label{sys1_2}
|\beta |^2 &+q(|\alpha |^2-|\beta |^2)= \\
& \qquad |\beta|^2 (|a|^4+|b|^4)+2 |\alpha|^2|a|^2|b|^2-2\chi, \nonumber
\end{align}
where $\chi=|\alpha| |\beta| (|a|^3|b| - |a||b|^3)\cos{\theta}$. Potential solutions of these equations are non-linear functions of the initial state, as implied by the no-go theorem in~\cite{allard2020no}. Moreover, for various $\alpha$, $\beta$, $a$ and $b$ there exist no solution $q\in [0,1]$ for  Eqs.~\eqref{sys1_1}-~\eqref{sys1_2} at all, see Appendix~\ref{app:Simple}. \\

However, we can model the memory flip, for arbitrary settings, if we assume different flipping probabilities for the different outcomes encoded in the memory register at $t_1$, see Fig.~\ref{Flip}(b). In that case we get the following two equations for $p(f_2)$
\begin{align}
\label{sys2_1}
|\alpha |^2 (1-q^0)&+|\beta |^2q^1=\\ 
& \qquad |\alpha|^2 (|a|^4+|b|^4)+2 |\beta|^2|a|^2|b|^2 +2\chi \nonumber \\
\label{sys2_2}
|\beta |^2 (1-q^1) &+|\alpha |^2q^0= \\
& \qquad |\beta|^2 (|a|^4+|b|^4)+2 |\alpha|^2|a|^2|b|^2-2\chi, \nonumber
\end{align}
which now always have solutions $q^0$ and $q^1$ $\in [0,1]$. Note that, now, these solutions are not necessarily unique and one can impose further conditions on them. For example, we could require that the memory flip is as symmetric as possible, i.e. $q^1=q^0+\epsilon$ with $|\epsilon|$ being minimal. This additional requirement, ensures that, whenever Eqs.~\eqref{sys1_1} -~\eqref{sys1_2} have a solutions we recovery those when solving Eqs.~\eqref{sys2_1} -~\eqref{sys2_2}. \\

In what follows we will consider whether the friend can be aware of these flipping probabilities. Note that, {such an awareness of changes to the memory register will necessarily create a tension with the complementarity principle. For example, if the friend} was aware of whether individual memory registers have been flipped or not, she could perfectly reconstruct the content of her memory before Wigner's measurement from the content after Wigner's measurement and her knowledge about the flip. {Together with the information about which result Wigner observed in his measurement observer and superobserver could then have complete information about the values of complementary observables in a single run}. One can, however, imagine scenarios where {the friend has a much more limited awareness of the changes to her memory. For example, let} the friend, after having stored the measurement results of multiple runs in different memory registers, be subject to Wigner performing his measurement on all these registers in parallel. The friend could then have \emph{limited} knowledge of how much her memory registers have been altered, without knowing whether individual registers have been flipped or not. This would correspond to a more direct perception of flipping probabilities $q^0$ and $q^1$. {In the next section we will apply an even weaker notion of the friend's awareness of changes to her memory, namely that she can only recognize whether the results stored in multiple memory registers have been flipped rather remained the same or not.} As we will show below, even such a perception of the changes of the friend's memory conflicts with the no-signaling principle.

\subsection{Memory changes in extended Wigner's-friend setups}
\label{Changes_E}

\begin{figure}[hbt!]
\begin{tikzpicture}[scale=0.57]

  \draw[<->,>=latex, thick,] (-2.5,0)--  (2.5,0);
  \node[fill=YellowGreen!30, draw=YellowGreen, thick] (s) at (0,0) {$\ket{\Phi}$};

  \draw[thick,] (5.5,1.5) rectangle (1.5,-1.5);
  \node[] (m) at (4,0) {\includegraphics[scale=0.15]{friend.jpg}};
    \node[] (F) at (4,-2) {F};
  \draw[->,>=latex,thick] (5.5,0) -- (6.75 ,0);
  \node[] (M) at (7.75,0) {\includegraphics[scale=0.05]{Wigner.jpg}};
    \node[] (W) at (7.75,-2) {W};
  \node[] (M) at (-4,0) {\includegraphics[scale=0.15]{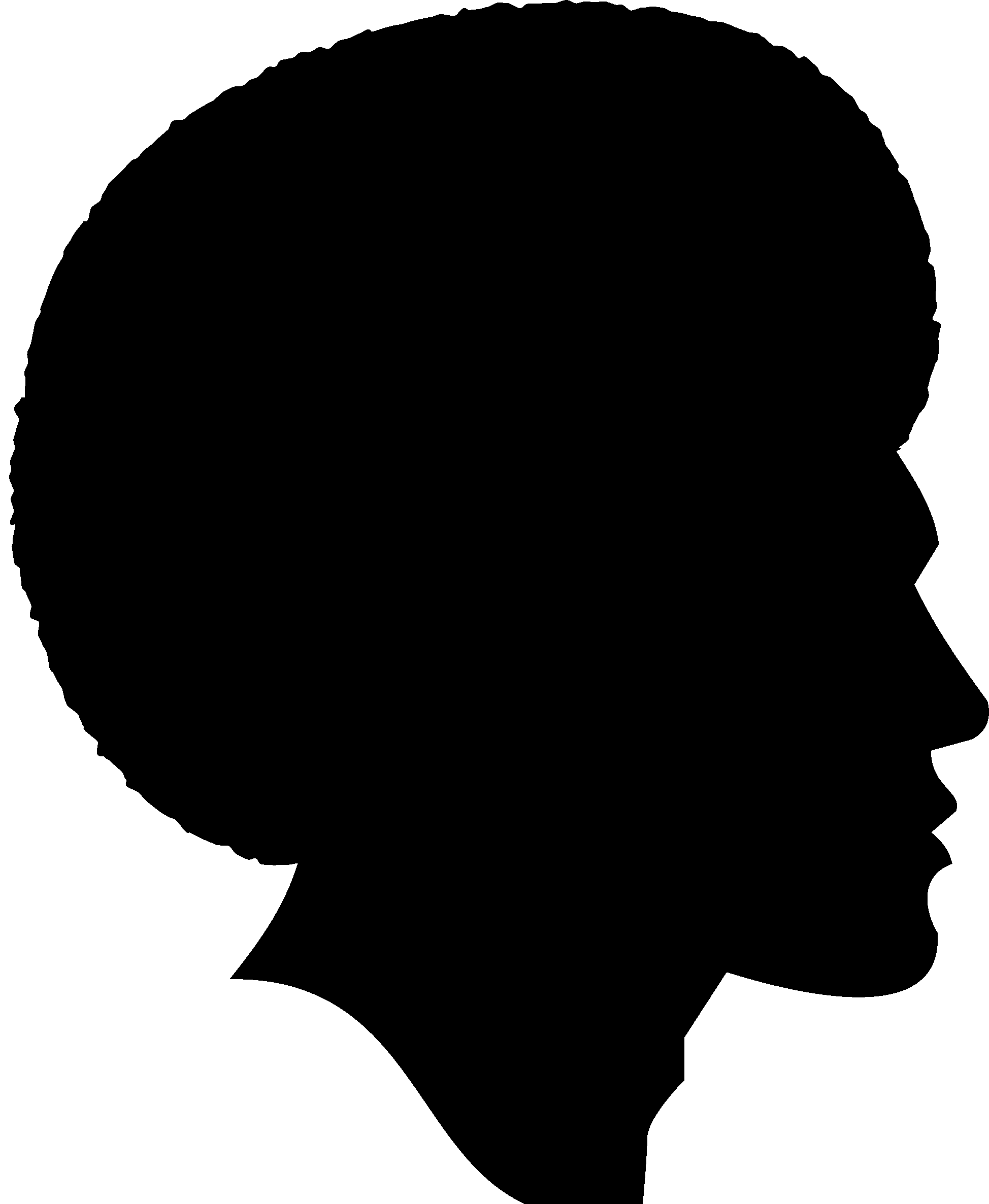}};
    \node[] (B) at (-4,-2) {B};
 \end{tikzpicture}
 \caption{The simplest version of an extended Wigner's friend experiment comprises a Wigner's friend setup and an additional observer Bob. The source emits a bipartite state, one part of which is measured by the friend in a fixed basis, while the other part is measured by Bob. Like in the original setup in Fig.~\ref{Simple Wigner}, Wigner measures the friend together with the respective subsystem, thereby potentially altering the result encoded in the friend's memory.
 }
    \label{Wigner_ext}
\end{figure}

We now consider extended Wigner's friend experiments, where the friend measures only part of a bipartite quantum state and, again, try to model the changes of her memory. The simplest version of an extended Wigner's friend experiment consists of one Wigner's friend setup and an additional observer Bob, who measures the second part of the bipartite initial state, see Fig.~\ref{Wigner_ext}. Consider the initial two qubit state $\ket{\Phi}= \alpha \ket{0,1} + \beta \ket{1,0}$. The first qubit is measured by the friend in the computational basis at time $t_F$. Bob performs a measurement corresponding to states $\ket{B=0}= \mu \ket{0}+\nu \ket{1}$ and $\ket{B=1}= \nu^* \ket{0}- \mu^* \ket{1}$ on the second qubit at some time $t_B>t_F$. Finally, at $t_W$, Wigner performs the  same measurement as in the simple Wigner's friend scenario in section~\ref{Changes}, given by states $\ket{W=1}$ and $\ket{W=2}$ and their orthogonal complement.\\

Similar to the simple Wigner's friend scenario, let us now consider the results stored in the friend's and Bob's memory across the time sequence $t_F<t_1<t_B<t_2<t_W<t_3$. The detailed calculations are given in Appendix~\ref{app:Ext}. After the friend performed her measurement the probabilities for the results $\bm 0$ and $\bm 1$ being stored in her memory are again simply
\begin{align}
\label{t1_0}
p(f_1=0)=& |\alpha|^2  \\
\label{t1_1}
p(f_1=1)=& |\beta|^2 . 
\end{align}
When performing his measurement, Bob will see results $B$ with the following probabilities.
\begin{align}
\label{t2_b0}
p(B_2=0)=& |\alpha|^2 |\nu|^2+|\beta|^2|\mu|^2  \\
\label{t2_b1}
p(B_2=1)=& |\alpha|^2 |\mu|^2+|\beta|^2|\nu|^2 ,  
\end{align}
where $B_2$ can be thought of as a result stored in Bob's memory at time $t_2$. The probabilities for the friend's memory stay the same  before and after Bob's measurement, i.e. $p(f_2)=p(f_1)$, as otherwise Bob could signal superluminaly to the other side. Until Wigner performs his measurement the friend and Bob are simply measuring two parts of a bipartite quantum state, nothing one of them does should change the perception of the other. %If the action of one were to alter the measured result of the other, this would heavily contradict our current empirical evidence. 
Finally, after Wigner's measurement, the probabilities for the result stored in the friend's memory are given by
\begin{align}
\label{t3_0}
p(f_3=0)=& |\alpha|^2 (|a|^4+|b|^4)+2|\beta|^2|a|^2|b|^2 \\
\label{t3_1}
p(f_3=1)=&  |\beta|^2 (|a|^4+|b|^4)+2|\alpha|^2|a|^2|b|^2 ,  
\end{align}
while those for Bob's result stay the same $p(B_3)=p(B_2)$. Moreover, the probabilities for the friend's memory entries are independent of Bob's measurement. These two facts, again, show that in terms of perceived measurement results the no-signaling condition is respected in the setup. This should not come as a surprise, since the respective probabilities are given by quantum theory, which is known to be non-signaling. \\

\begin{table}
\setlength{\tabcolsep}{12pt}
\renewcommand{\arraystretch}{1.25}
\resizebox{0.37\textwidth}{!}{%
\begin{tabular}{c|c|c}
\multicolumn{3}{c}{
$p(f,B) \text{ for }\bm{t_2}<t_W$
}  \\ \hline \hline
\diagbox[innerwidth=20pt]{$f_2$}{$B_2$} & 0 & 1 \\ \hline 
0 & $|\alpha|^2 |\nu|^2$ &$|\alpha|^2 |\mu|^2$  \\ \hline
1 & $|\beta|^2|\mu|^2 $ & $|\beta|^2|\nu|^2$  \\ \hline
\multicolumn{3}{c}{}
\end{tabular}
}
\resizebox{0.49\textwidth}{!}{%
\begin{tabular}{c|c|c}
\multicolumn{3}{c}{
\large{$p(f,B) \text{ for } \bm{t_3}>t_W$}
}  \\ \hline \hline
\diagbox[innerwidth=20pt]{$f_3$}{$B_3$} & 0 & 1 \\ \hline 
\multirow{2}*{0} & $ |\alpha|^2 |\nu|^2 (|a|^4+|b|^4)+$ &$|\alpha|^2 |\mu|^2 (|a|^4+|b|^4)+$  \\
 	&$2 |\beta|^2 |\mu|^2|a|^2|b|^2 +2\xi$ & $2 |\beta|^2 |\nu|^2|a|^2|b|^2 - 2\xi$ \\ \hline
\multirow{2}*{1} & $|\beta|^2 |\mu|^2 (|a|^4+|b|^4)+$ &  $ |\beta|^2 |\nu|^2 (|a|^4+|b|^4)+$ \\
	& $2 |\alpha|^2 |\nu|^2|a|^2|b|^2 - 2\xi$ & $2 |\alpha|^2 |\mu|^2|a|^2|b|^2 +2\xi$ \\ \hline
\multicolumn{3}{c}{}\\
\multicolumn{3}{c}{ with $\xi=(|a|^3|b|-|a||b|^3)|\alpha||\beta||\mu||\nu| \cos \vartheta$}\\
\end{tabular}%
}
\caption{The joint probabilities for the results $f$ and $B$ recorded in the friend's and Bob's memory respectively. The top table shows the joint probability before Wigner's measurement. The bottom table presents the joint probability after Wigner has performed his measurement. The detailed calculation and the explicit form of relative phase factor $\vartheta$ can be found in Appendix~\ref{app:Ext}.}
\label{JointProb}
\end{table}

When trying to consistently model the change of the friend's memory, we further have to consider the joint probabilities for the results of the friend and Bob, see Tab.~\ref{JointProb}. The joint probabilities at $t_3$ and $t_2$ are related as follows
\begin{align}
\label{Joint_change1}
p(f_3,B_3)=\sum_{f_2,B_2} p(f_2,B_2)p(f_3,B_3|f_2,B_2) 
\end{align}
where the conditional probability $p(f_3,B_3|f_2,B_2)$, which still needs to be specified, encodes how the records of the results of the friend and Bob change from before to after Wigner's measurement. Note that, these probabilities are not given by quantum theory, i.e. they \emph{cannot} be written as $\tr{O(t_2,t_3)\proj{\Psi}}$ for some two-time observable $O(t_2,t_3)$ and state $\ket{\Psi}$. The conditional probability in Eq.~\eqref{Joint_change1} can further be written as $p(f_3,B_3|f_2,B_2)=p(B_3|f_2,B_2,f_3)p(f_3|f_2,B_2)$. Since the result stored in Bob's memory should not be affected by any action of Wigner between $t_2$ and $t_3$, i.e. $p(B_3|B_2)=\delta_{B_3B_2}$, one has that 
\begin{align}
\label{Joint_change2}
p(f_3,B_3)=\sum_{f_2,B_2} p(f_2,B_2) p(f_3|f_2,B_2) \delta_{B_3B_2}. 
\end{align}
In order to make the joint probabilities at different times empirically accessible, imagine that both the friend and Bob have multiple memory registers available, such that they can record results for multiple rounds of measurements. Wigner can then measure all the friend's registers and respective subsystems in his computational basis (i.e. $\ket{0}_S\ket{0}_F$, $\ket{1}_S\ket{1}_F$), which returns the probabilities with which the friend observed one or the other result during her measurement. In this case, Bob, the friend and Wigner will establish the joint probability before Wigner's measurement $p(f_2,B_2)$. In another set of rounds, however, Wigner will first measure all the registers of his friend and the respective subsystems in some other basis (i.e. $\ket{W=1}_{SF}$, $\ket{W=2}_{SF}$ and their orthogonal complement) and perform the computational basis measurement afterwards. In this case $p(f_3,B_3)\neq p(f_2,B_2)$ will be established. \\

Similar to section~\ref{Changes} we try to model the change of the friend's memory by flipping probabilities $q^0$ and $q^1$ and obtain the following set of equations:
\begin{align}
\label{joint00}
&(1-q^0)|\alpha|^2 |\nu|^2+q^1|\beta|^2|\mu|^2=  p(0_3,0_3)\\
	%&\qquad \qquad |\alpha|^2 |\nu|^2 (|a|^4+|b|^4)+ 2 |\beta|^2 |\mu|^2|a|^2|b|^2 +2\xi\nonumber \\
\label{joint01}
&(1-q^0)|\alpha|^2 |\mu|^2+q^1|\beta|^2|\nu|^2= p(0_3,1_3)\\
 	%&\qquad \qquad |\alpha|^2 |\mu|^2 (|a|^4+|b|^4)+ 2 |\beta|^2 |\nu|^2|a|^2|b|^2 - 2\xi \nonumber \\
 \label{joint10}
&(1-q^1)|\beta|^2|\mu|^2 +q^0|\alpha|^2 |\nu|^2 =p(1_3,0_3)\\
	%&\qquad \qquad |\beta|^2 |\mu|^2 (|a|^4+|b|^4)+2 |\alpha|^2 |\nu|^2|a|^2|b|^2 - 2\xi \nonumber \\
\label{joint11}
 &(1-q^1)|\beta|^2|\nu|^2+q^0|\alpha|^2 |\mu|^2 =p(1_3,1_3),
 	%& \qquad \qquad |\beta|^2 |\nu|^2 (|a|^4+|b|^4)+ 2 |\alpha|^2 |\mu|^2|a|^2|b|^2 +2\xi,\nonumber   
\end{align}
where the $p(f_3,B_3)$ are given by the bottom table of Tab.\ref{JointProb}. This set of equations does not always have solutions $q^0,q^1 \in [0,1]$. But we can introduce a dependence on Bob's result to increase the number of parameters to $q^{00},q^{01},q^{10}$ and $q^{11}$ to model the memory change for all possible settings. As in the simple Wigner's friend scenario, this introduces non-uniqueness for the potential solutions. Most importantly, however, solutions of Eqs.~\eqref{joint00} -~\eqref{joint11} and their modifications will depend non-trivially on $\mu$ and $\nu$ and, therefore, on Bob's \emph{measurement setting}, see Appendix~\ref{app:Ext} for details. This means the flipping probabilities for the results stored in the friend's memory must, in general, be signaling. As we show in the next section, this could be used to transmit information from Bob to the friend and Wigner at superluminal speed, even if the friend has only a very limited awareness of the change of her memory. \\

The possibility of superluminal signaling can be understood in the following way. The result stored in the friend's memory before Wigner's measurement plays the role of a hidden variable, that encodes the result Wigner would have obtained if he had measured in the computational basis. In order to be compatible with the violations of Bell-like inequalities from the no-go theorems in~\cite{Bong2020,brukner2018no}, these variables must change non-locally. In other words, during Wigner's measurement the records in the friend's memory will change in non-local dependence of the choice of Bob's measurement setting. If the friend was now aware of the change of her memory, even just partially, this would constitute superluminal signaling from Bob to her. Moreover, she could inform Wigner about the change after all measurements have been performed, enabling superluminal signaling also between Bob and Wigner.

\section{A signaling protocol}
\label{ConfSig}

We now describe a specific hypothetical protocol for the extended Wigner's friend scenario depicted in Fig.~\ref{Wigner_ext} and show that it could be used to signal at superluminal speed if the friend was aware of the change of her memory. We assume that the flipping probabilities are such that the dependence on Bob's outcome is minimal, meaning that we obtain the solutions to the simpler Eqs.~\eqref{joint00}-~\eqref{joint11}, if they exist. \footnote{Note, that this assumption simplifies the calculation for the example we present here, but is not necessary for establishing a signaling protocol in general.} \\

Consider, a version of the setup in Fig.~\ref{Wigner_ext} where Bob resides in spacelike separation from Wigner and his friend. Let Bob and the friend have multiple memory registers available and perform a sequence of $N\gg1$ measurements. Then Wigner performs his measurement, given by states $\ket{W=1}$, $\ket{W=2}$ which are superpositions of $\ket{0}_S\ket{0}_F$ and $\ket{1}_S\ket{1}_F$, and their orthogonal complement, on the friend's $N$ memory registers and the respective qubits. Now suppose they repeat the sequence of  $N$ measurements multiple times, while Bob chooses between two different measurement settings for these repetitions. Within each repetition, however, he performs the same chosen measurement all $N$ times he receives his subsystem from the source. We further assume that the friend has some notion on how much her collection of $N$ memory registers has been changed by Wigner's measurement. For example, the friend could be aware of whether the results stored in her $N$ memory registers have been flipped rather than stayed the same during Wigner's measurement, i.e. whether the flipping probability is larger or smaller than $1/2$. \\

Now, let the source emit $N$ maximally entangled two-qubit state
\begin{align}
\label{initial_state}
\ket{\Phi}=\frac{1}{\sqrt{2}}\left( \ket{0}\ket{1}+\ket{1}\ket{0} \right),
\end{align}
and Bob choose between measuring his $N$ qubits either in the computational basis $\ket{0}$, $\ket{1}$ or in the basis $\ket{B=0}=1/\sqrt{3}\ket{0}+\sqrt{2/3}\ket{1}$, $\ket{B=1}=\sqrt{2/3}\ket{0} - 1/\sqrt{3}\ket{1}$. Wigner always performs the measurement given by the states
\begin{align}
\label{Wigner_msm}
\ket{W=1}&= \sin\left( \frac{\pi}{8}\right)\ket{0,0}_{SF}+  \cos\left( \frac{\pi}{8}\right) \ket{1,1}_{SF} \\
\ket{W=2}&= \cos\left( \frac{\pi}{8}\right) \ket{0,0}_{SF} -  \sin\left( \frac{\pi}{8}\right)  \ket{1,1}_{SF},
\end{align}
and the orthogonal complement. If Bob measures his qubit in the computational basis, the joint probabilities before Wigner's measurement are 
\begin{equation*}
\renewcommand{\arraystretch}{1.25}
  \begin{array}{c|c|c}
  p(f_2,B_2)  & B=0 & B=1 \\  [0.25em]  \hline
  f= 0 & 0 & \frac{1}{2}  \\[0.25em]
   f= 1 &  \frac{1}{2} & 0 \;,
  \end{array}
\end{equation*}
while after Wigner's measurement, we obtain
\begin{equation*}
\renewcommand{\arraystretch}{1.25}
  \begin{array}{c|c|c}
  p(f_3,B_3)  & B=0 & B=1 \\  [0.25em]  \hline
  f= 0 & \frac{1}{8} & \frac{3}{8}  \\[0.25em]
   f= 1 &  \frac{3}{8} & \frac{1}{8} \;.
  \end{array}
\end{equation*}
In this case, the solution to Eqs.~\eqref{joint00}-~\eqref{joint11} is $q^0=q^1=q=1/4$, which means that the result stored in the friend's memory rather stays the same (with probability $1-q=3/4$) than gets flipped during Wigner's measurement. On the other hand, if Bob measures his $N$ qubits in the basis $\ket{B=0}=1/\sqrt{3}\ket{0}+\sqrt{2/3}\ket{1}$, $\ket{B=1}=\sqrt{2/3}\ket{0} - 1/\sqrt{3}\ket{1}$, the joint probabilities before Wigner measures are
\begin{equation*}
\renewcommand{\arraystretch}{1.25}
  \begin{array}{c|c|c}
  p(f_2,B_2)  & B=0 & B=1 \\  [0.25em]  \hline
  f= 0 & \frac{1}{3} & \frac{1}{6}  \\[0.25em]
   f= 1 &  \frac{1}{6} & \frac{1}{3} \;,
  \end{array}
\end{equation*}
while after Wigner's measurement we have
\begin{equation*}
\renewcommand{\arraystretch}{1.25}
  \begin{array}{c|c|c}
  p(f_3,B_3)  & B=0 & B=1 \\  [0.25em]  \hline
  f= 0 & \frac{1}{24} \left(7-2\sqrt{2}\right)
 & \frac{1}{24} \left(5+2\sqrt{2}\right) \\[0.25em]
   f= 1 &  \frac{1}{24} \left(5+2\sqrt{2}\right)& \frac{1}{24} \left(7-2\sqrt{2}\right) \;,
  \end{array}
\end{equation*}
and the solution to Eqs.~\eqref{joint00}-~\eqref{joint11} is now $q^0=q^1=q=1/4+ 1/\sqrt{2}\approx 0.96$, which means that the friend's memory registers get flipped more often than they stay the same when Wigner performs his measurement. Hence, even with the very limited awareness of whether her collection of $N$ memory registers got flipped rather than not in a repetition of the sequence, the friend could distinguish between these two measurement settings of Bob and communicate this to Wigner. Hence, Bob could, via his choice of measurement setting for each repetition, transmit information to Wigner and his friend at superluminal speed provided that Wigner's collective measurement happens outside of the future light cone of Bob's measurements. In turn, under the assumption of the no-signaling condition, no such perception of the change in the friend's memory in Wigner's friend experiments is possible. In other words, the friend's inner thoughts must be changed by Wigner's measurement such that no knowledge about the memory before Wigner's measurement remains. This is in agreement with the assumptions in Ref.~\cite{wiseman2023thoughtful}.

\section{Conclusions}
\label{Conclusions}

We considered the compatibility of certain models of the observer's (the friend's) internal thoughts in Wigner's friend scenarios with the no-signaling condition. In general, Wigner's measurement affects the record of the measurement result stored in the friend's memory register. This change in perception, just like the result the friend initially observed, is not directly accessible to Wigner and does not correspond to any observable on the joint system of S+F. However, identifying the friend with states of perception and inner thoughts begs the question in how far she can be aware of what happens to her memory. We considered the idea that the friend might have \emph{some} perception of the change Wigner's measurement induces to her memory. More concretely, we described the case where the friend can have multiple registers available for storing results which are then each measured by Wigner in some fixed basis. The friend could have a notion of the overall change to these memory registers (for example, the probability of entries being flipped during Wigner's measurement) without having a persistent perception of individual results across Wigner's measurement.
We modeled the change of the friend's memory and found that, even if the friend has only a very limited awareness of the change of her memory, this awareness could be used by a distant observer (Bob) to signal superluminalny to the friend and Wigner. While this possibility to signal is general, there is still considerable freedom in how to model the memory change, i.e. the flipping probabilities. How to further determine these flipping probabilities in a Winger's friend setup by physically well motivated additional constraints, is an interesting question. It is, however, beyond the scope of this paper and we leave it for future work. We take the conflict with the no-signaling principle demonstrated in this work as a strong argument against the friend being able to have such an awareness, meaning that not only can the friend not know about the specific result she observed before Wigner's measurement once her memory changed, she cannot have awareness of the change either. 

\section*{Acknowledgements}
We want to thank Tobias Sutter for much appreciated feedback. This research was funded in whole or in part by the Austrian Science Fund (FWF) 10.55776/COE1 (Quantum Science Austria), [10.55776/F71] (BeyondC), [10.55776/RG3] (Reseacrh Group 3) and [ESP520] (ESPRIT). For Open Access purposes, the authors have applied a CC BY public copyright license to any author accepted manuscript version arising from this submission. This research was supported by FQXi FFF Grant number FQXi-RFP-CPW-2005 from the Foundational Questions Institute and Fetzer Franklin Fund, a donor advised fund of  the Silicon Valley Community Foundation as well as the financial support of ID61466 and ID62312 grants from the John Templeton Foundation as part of The Quantum Information Structure of Spacetime (QISS) Project (qiss.fr). The opinions expressed in this publication are those of the authors and do not necessarily reflect the views of the John Templeton Foundation.
\newpage

\bibliography{F-mem}
\bibliographystyle{quantum}

\newpage

%%%%%%%%%%%%%%%%%%%%%%%%%%%%%%%%%%%%%%%%%%%%%%%%%%%%%
\begin{appendix}

\section{Memory changes in the simplest Wigner's friend setup}
\label{app:Simple}

Assuming that a unitary description of a setup is valid even if the total system involves observers, we have the following state evolution for the simplest Wigner's friend experiment, depicted in Fig.~\ref{Simple Wigner}. Initially the state is given by
\begin{align}
\label{Psi0s}
\ket{\Psi(t_0)}=\left( \alpha \ket{0}_S+\beta \ket{1}_S\right)\ket{r}_F\ket{r}_W,
\end{align}
where $\ket{r}$ denotes the ready state of an observer, when he or she has not performed a measurement yet. Measurements are described by entangling unitaries which correlate the respective observer (F and W) with the measured quantum system
\begin{align}
\label{msm}
U_O: \ket{i}_S\ket{r}_O \mapsto  \ket{i}_S\ket{I_i}_O \qquad \forall i,
\end{align}
where $\ket{i}$ are the eigenstates of the observable being measured and $\ket{I_i}$ is the state of the observer having registered outcome $i$. After the friend measured the system S in the computational basis, we therefore obtain the state
\begin{align}
\label{Psi1s}
\ket{\Psi(t_1)}=\left( \alpha \ket{0,\bm 0}_{SF}+\beta \ket{1, \bm 1}_{SF}\right)\ket{r}_W.
\end{align}
After Wigner performed his measurement with eigenvectors
\begin{align}
\ket{W=1}&= a\ket{0}_S\ket{\bm0}_F+b\ket{1}_S\ket{\bm1}_F\\
\ket{W=2}&=b^*\ket{0}_S\ket{\bm0}_F-a^*\ket{1}_S\ket{\bm1}_F,
\end{align}
we obtain the overall state
\begin{align}
\label{Psi2s}
\ket{\Psi(t_2)}&= a(\alpha a^*+\beta b^*)\ket{0,\bm 0}_{SF}\ket{\bm1}_W \nonumber \\
	&+ b(\alpha a^*+\beta b^*)\ket{1,\bm 1}_{SF}\ket{\bm1}_W \nonumber \\
	&+b^*(\alpha b -  \beta a)\ket{0,\bm 0}_{SF}\ket{\bm2}_W \nonumber \\
	&-a^*(\alpha b -  \beta a)\ket{1,\bm 1}_{SF}\ket{\bm 2}_W.
\end{align}
The probability of an observer having registered outcome $x$ at time $t_j$ is then given by
\begin{align}
\label{prob}
p(x_j)= \tr \left( \mathds{1}\otimes\proj{X_x}_O \cdot \proj{\Psi(t_j)} \right)
\end{align}
Writing the complex coefficients as $z=|z|e^{i\phi_z}$ we obtain the following probabilities for the result encoded in the friend's state $\ket{\cdot}_F$, i.e. the entry in her memory register. Before Wigner's measurement we have
\begin{align}
p(0_1)&=\tr \left(\mathds{1}\otimes \proj{\bm 0}_F \cdot \proj{\Psi(t_1)} \right)= |\alpha|^2 \\
p(1_1)&=\tr \left( \mathds{1}\otimes \proj{\bm 1}_F \cdot \proj{\Psi(t_1)} \right)= |\beta|^2 ,
\end{align}
and after Wigner's measurement we obtain
\begin{align}
p(0_2)=&\tr \left(\mathds{1}\otimes \proj{\bm 0}_F \cdot \proj{\Psi(t_2)} \right)= \nonumber \\
& |\alpha|^2 (|a|^4+|b|^4)+2 |\beta|^2|a|^2|b|^2 \nonumber\\
\label{p0_2sim}
	&+2|\alpha| |\beta| (|a|^3|b| - |a||b|^3)\cos{\theta} 
\end{align}	
\begin{align}
p(1_2)=&\tr \left(\mathds{1}\otimes \proj{\bm 1}_F \cdot \proj{\Psi(t_2)} \right)= \nonumber \\
&|\beta|^2 (|a|^4+|b|^4)+2 |\alpha|^2|a|^2|b|^2 \nonumber \\
\label{p1_2sim}
	&-2|\alpha| |\beta| (|a|^3|b| - |a||b|^3)\cos{\theta},
\end{align}
where $\theta=\phi_{\alpha}-\phi_{\beta}+\phi_b -\phi_a$ and we define $|\alpha| |\beta| (|a|^3|b| - |a||b|^3)\cos{\theta}=:\chi$. Modeling the change of the memory by flipping probability $q$ we can further write
\begin{align}
\nonumber
p(0_2)&=(1-q)p(0_1) + qp(1_1)\\
&=(1-q)|\alpha|^2 + q|\beta|^2\\
\nonumber
p(1_2)&= qp(0_1) + (1-q)p(1_1)\\
&=q|\alpha|^2 + (1-q)|\beta|^2,
\end{align}
which, when equated with~\eqref{p0_2sim} and~\eqref{p1_2sim}, gives Eqs.~\eqref{sys1_1} -~\eqref{sys1_2} in the main text. Adding and subtracting the latter leads to
\begin{align*}
&|\alpha|^2+|\beta|^2=|a|^4+|b|^4 + 2|a|^2|b|^2\\
&(|\alpha|^2-|\beta|^2)(1-2q)= (|\alpha|^2-|\beta|^2)[|a|^2- |b|^2]^2+4\chi,
\end{align*}
from which it is easy to see that there is not always a solution $q$. For example, if $|\alpha|=|\beta|=1/\sqrt{2}$ solutions $q$ only exist for $|a|=|b|=1/\sqrt{2}$ (where $q=1/2$) and $a=0,b=1$ or $a=1,b=0$ (where $q=0$). \\

One way to obtain solutions for all settings is to increase the number of variables in Eqs.~\eqref{sys1_1} -~\eqref{sys1_2} by assuming different flipping probabilities depending on which outcome the friend initially observed, see Fig.~\ref{Flip}(b). In that case we obtain
\begin{align}
p(0_2)=&(1-q^0)|\alpha|^2 + q^1|\beta|^2\\
p(1_2)=&q^0|\alpha|^2 + (1-q^1)|\beta|^2,
\end{align}
which then leads to Eqs.~\eqref{sys2_1} -~\eqref{sys2_2} in the main text. These, in turn, give equations
\begin{align*}
&|\alpha|^2+|\beta|^2=1\\
&|\alpha|^2(1-2q^0)-|\beta|^2(1-2q^1)\\
& \qquad \qquad \qquad= (|\alpha|^2-|\beta|^2)[|a|^2- |b|^2]^2+4\chi,
\end{align*}
which now can be solved for $q^0,q^1$. Note that, the solutions are not unique. For example, if $|\alpha|=|\beta|=1/\sqrt{2}$ and further $|a|=|b|=1/\sqrt{2}$, we obtain only $q^0-q^1=0$ from Eqs.~\eqref{sys2_1} -~\eqref{sys2_2}, which is solved by any number $q^0=q^1\in[0,1]$. One can further require that the solutions are as symmetric as possible $q^1=q^0+\epsilon$ with minimal $|\epsilon|$, which means that if there is a solution to Eqs.~\eqref{sys1_1} -~\eqref{sys1_2} we recover it when solving Eqs.~\eqref{sys2_1} -~\eqref{sys2_2}.

\section{Memory changes in an extended Wigner's friend experiment}
\label{app:Ext}

Similar to section~\ref{app:Simple}, we now consider the extended Wigner's friend experiment in Fig.~\ref{Wigner_ext}. Before any measurement happened we have the overall state.
\begin{align}
\label{Psi0e}
\ket{\Psi(t_0)}= &\left( \alpha \ket{0,1}_{12} +\beta \ket{1,0}_{12}\right)\ket{r}_F\ket{r}_B\ket{r}_W,
\end{align}
where, again $\ket{r}$ denotes the ready state of an observer. After the friend performed her measurement in the computational basis, the state becomes
\begin{align}
\label{Psi1e}
\ket{\Psi(t_1)}=&\Big( \alpha \ket{0,\bm0}_{1F}\ket{1}_{2}  \\ \nonumber 
& +\beta \ket{1,\bm1}_{1F}\ket{0}_{2}\Big)\ket{r}_B\ket{r}_W.
\end{align}
Then Bob measures the other subsystem in some basis, $\ket{B=0}= \mu \ket{0}+\nu \ket{1}$, $\ket{B=1}= \nu^* \ket{0}- \mu^* \ket{1}$ leading to overall state
\begin{align}
\label{Psi2e}
\ket{\Psi(t_2)}=&\Big( \alpha \nu^* \ket{0,\bm0}_{1F}\ket{0, B=0}_{2B} \\ \nonumber
& -\alpha \mu \ket{0,\bm0}_{1F}\ket{1, B=1}_{2B} \\ \nonumber
& +\beta \mu^* \ket{1,\bm1}_{1F}\ket{0, B=0}_{2B}\\ \nonumber
&  +\beta \nu \ket{1,\bm1}_{1F}\ket{1, B=1}_{2B} \Big)\ket{r}_W.
\end{align}
Finally, after Wigner's measurement of $\ket{W=1}= a\ket{0,0}_{1F}+b\ket{1,1}_{1F}$ and $\ket{W=2}=b^*\ket{0,0}_{1F}-a^*\ket{1,1}_{1F}$ the unitarily evolved state is 
\begin{align}
\label{Psi3e}
&\ket{\Psi(t_3)}= \nonumber \\
&\qquad (\alpha \nu^* a^*+\beta\mu^* b^*)  \ket{W=1,\bm 1}_{1F,W}\ket{B=0,0}_{2,B} \nonumber \\
&\qquad +(\beta \nu b^*-\alpha\mu a^*) \ket{W=1,\bm1}_{1F,W}\ket{B=1,1}_{2,B}  \nonumber \\
&\qquad +(\alpha \nu^* b-\beta\mu^* a) \ket{W=2,\bm2}_{1F,W}\ket{B=0,0}_{2,B} \nonumber \\
&\qquad -(\alpha \mu b+\beta \nu a) \ket{W=2,\bm2}_{1F,W}\ket{B=1,1}_{2,B}.
\end{align}
From Eqs.~\eqref{Psi1e}-\eqref{Psi3e} we can calculate the individual probabilities for the friend and Bob using Eq.~\eqref{prob} both before and after Wigner's measurement. Before Wigner measures, the  probabilities for the results stored in the friend's memory are
\begin{align}
\label{ft_1_0}
p(f_1=0)=& |\alpha|^2=p(f_2=0)  \\
\label{ft1_1}
p(f_1=1)=& |\beta|^2=p(f_2=0) ,  
\end{align}
which do not change upon Bob's measurement, as was to be expected since, before Wigner's measurement, the friend and Bob constitute two regular observers measuring different parts of a bipartite quantum state. Nothing one of them does should affect the result the other one observes.
For Bob's results we obtain
\begin{align}
\label{bt2_0}
p(B_2=0)=& |\alpha|^2 |\nu|^2+|\beta|^2|\mu|^2  \\
\label{bt2_1}
p(B_2=1)=& |\alpha|^2 |\mu|^2+|\beta|^2|\nu|^2 . 
\end{align}
The joint probabilities for two outcomes $x,y$ being stored at time $t_j$ in the memories of two observers $O_1$ and $O_2$ respectively is 
\begin{align}
\label{jprob}
p(x_j,y_j)= \tr \left( \proj{X}_{O_1} \otimes\proj{Y}_{O_2}  \cdot \proj{\Psi(t_j)} \right),
\end{align}
which gives the following joint probabilities for the results recoded by the friend and Bob before Wigner's measurement
\begin{equation}
\label{f2b2}
\renewcommand{\arraystretch}{1.25}
  \begin{array}{c|c|c}
  p(f_2,B_2)  & B=0 & B=1 \\  [0.25em]  \hline
  f= 0 & |\alpha|^2 |\nu|^2 & |\alpha|^2 |\mu|^2  \\[0.25em]
   f= 1 & |\beta|^2|\mu|^2 & |\beta|^2|\nu|^2 \;.
  \end{array}
\end{equation}
\begin{figure}
\includegraphics[width=0.49\textwidth]{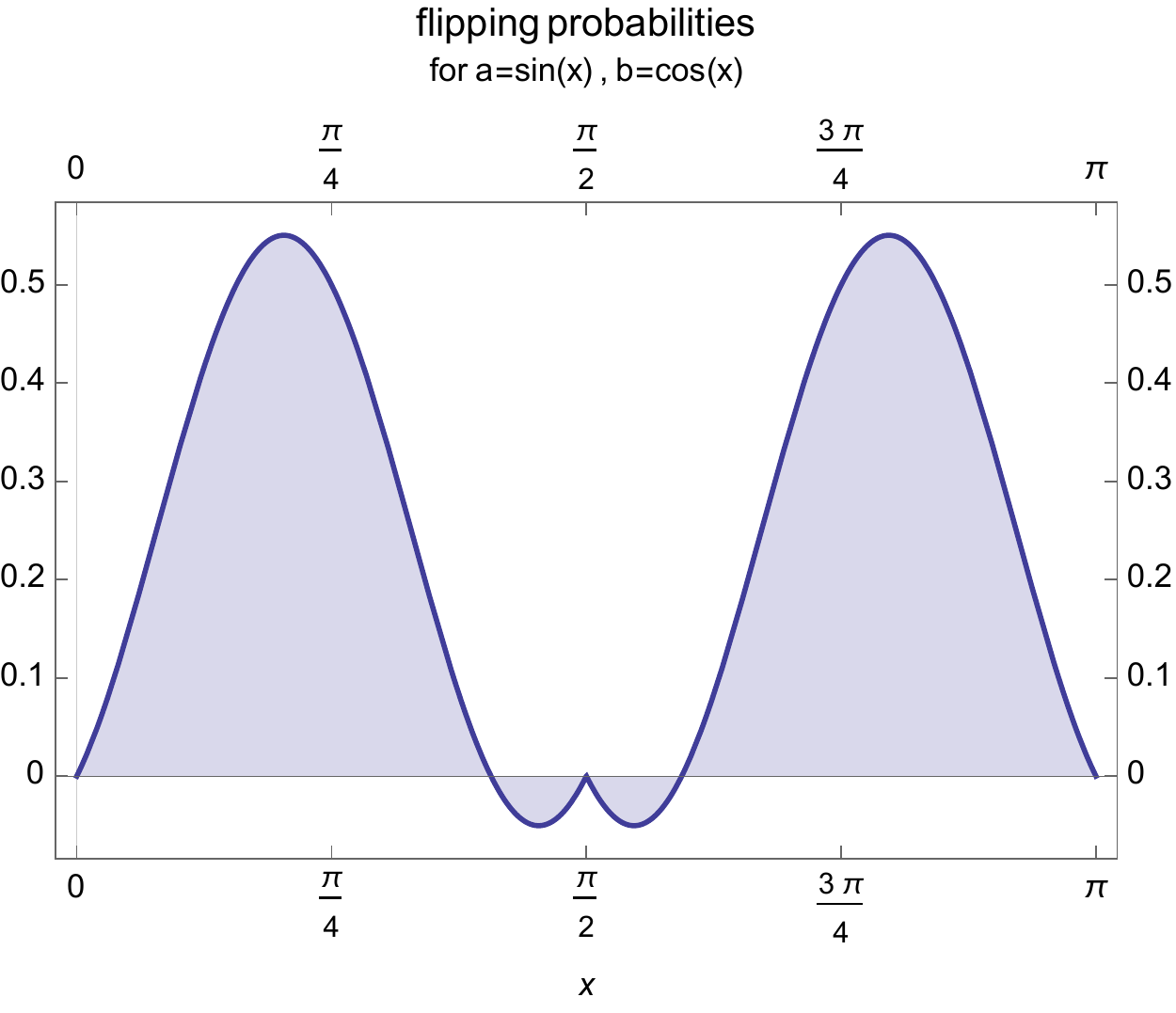}
\caption{Flipping probabilities $q^{00}=q^{11}$ (according to Eq.\eqref{Condi}) for an extended Wigner's friend setup with initial state $\ket{\Phi}=1/\sqrt{2}\left( \ket{0}\ket{1}+\ket{1}\ket{0} \right)$. Bob measures in the basis $\ket{B=0}=1/\sqrt{3}\ket{0}+\sqrt{2/3}\ket{1}$, $\ket{B=1}=\sqrt{2/3}\ket{0} - 1/\sqrt{3}\ket{1}$ and Wigner's measurement is given by $a=\sin(x)$, $b=\cos(x)$ and $\cos(\Delta\phi)=1$. Since these $q^{00}=q^{11}$ become negative for some values of $x$, they are not always well defined probabilities.}
\label{signaling}
\end{figure}
After Wigner's measurement we obtain the following individual probabilities
\begin{align}
\label{ft_1_0}
p(f_3=0)=&|\alpha|^2 (|a|^4+|b|^4)+ 2 |\beta|^2 |a|^2|b|^2   \\
\label{ft1_1}
p(f_3=1)=& |\beta|^2 (|a|^4+|b|^4)+2 |\alpha|^2 |a|^2|b|^2 ,  
\end{align}
for the result stored in the friend's memory register and 
\begin{align}
\label{bt3_0}
p(B_3=0)=& |\alpha|^2 |\nu|^2+|\beta|^2|\mu|^2  =p(B_2=0)\\
\label{bt3_1}
p(B_3=1)=& |\alpha|^2 |\mu|^2+|\beta|^2|\nu|^2  =p(B_2=1)
\end{align}
for Bob's memory. The latter is the same as before Wigner's measurement. Since Bob is a regular observer, whose memory is not subject to any further measurement and who can in principle freely communicate to Wigner, his record of an observed result should not be affected by anything Wigner does to his friend.

The joint probabilities for the results stored in the friend's and Bob's memory after Wigner's measurement, according to Eq.~\eqref{jprob}, are given by
\begin{widetext}
\begin{equation*}
\renewcommand{\arraystretch}{1.25}
  \begin{array}{c|c| c}
  p(f_3,B_3)  & B=0  & B=1\\  [0.25em]  \hline
  f= 0 & |\alpha|^2 |\nu|^2 (|a|^4+|b|^4)+ 2 |\beta|^2 |\mu|^2|a|^2|b|^2 +2\xi  
   &  |\alpha|^2 |\mu|^2 (|a|^4+|b|^4)+ 2 |\beta|^2 |\nu|^2|a|^2|b|^2 - 2\xi  \\[0.25em]
   f= 1 &  |\beta|^2 |\mu|^2 (|a|^4+|b|^4)+2 |\alpha|^2 |\nu|^2|a|^2|b|^2 - 2\xi 
   &  |\beta|^2 |\nu|^2 (|a|^4+|b|^4)+ 2 |\alpha|^2 |\mu|^2|a|^2|b|^2 +2\xi \;.
  \end{array}
\end{equation*}
\centering with  $\xi=(|a|^3|b|-|a||b|^3)|\alpha||\beta||\mu||\nu| \cos (\phi_{\alpha}-\phi_{\beta}+\phi_{\mu}-\phi_{\nu}+\phi_b-\phi_a).$ 
\end{widetext}
We can now try to model the change in the friend's memory like we did in App.~\ref{app:Simple}, using different flipping probabilities depending on the friends initial result, and obtain the following system of equations\\
\begin{align}
\label{Ajoint00}
&(1-q^0)|\alpha|^2 |\nu|^2+q^1|\beta|^2|\mu|^2= \\
	&\qquad \qquad |\alpha|^2 |\nu|^2 (|a|^4+|b|^4)+ 2 |\beta|^2 |\mu|^2|a|^2|b|^2 +2\xi\nonumber \\
\label{Ajoint01}
&(1-q^0)|\alpha|^2 |\mu|^2+q^1|\beta|^2|\nu|^2= \\
 	&\qquad \qquad |\alpha|^2 |\mu|^2 (|a|^4+|b|^4)+ 2 |\beta|^2 |\nu|^2|a|^2|b|^2 - 2\xi \nonumber \\
 \label{Ajoint10}
&(1-q^1)|\beta|^2|\mu|^2 +q^0|\alpha|^2 |\nu|^2 =\\
	&\qquad \qquad |\beta|^2 |\mu|^2 (|a|^4+|b|^4)+2 |\alpha|^2 |\nu|^2|a|^2|b|^2 - 2\xi \nonumber \\
\label{Ajoint11}
 &(1-q^1)|\beta|^2|\nu|^2+q^0|\alpha|^2 |\mu|^2 =\\
 	& \qquad \qquad |\beta|^2 |\nu|^2 (|a|^4+|b|^4)+ 2 |\alpha|^2 |\mu|^2|a|^2|b|^2 +2\xi,\nonumber   
\end{align}
which, similar to before, do not  always have a solution. Consider, for example, the case where the source emits the maximally entangled state in Eq.~\eqref{initial_state} in the main text, Bob measures in the X-basis, i.e. $\mu= \nu =1/\sqrt{2}$, and Wigner's measurement contains the Bell-states $\ket{\phi^{\pm}}$, i.e. $\alpha=\beta=1/\sqrt{2}$. In this case Eqs.~\eqref{Ajoint00}-~\eqref{Ajoint11} become
\begin{align}
\nonumber
&\frac{1}{4} \left( 1-q^0+q^1\right)= \frac{1}{4}+\frac{1}{2}(|a|^3 |b|-|a| |b|^3)\cos(\Delta\phi) \\
\nonumber
&\frac{1}{4} \left( 1-q^0+q^1\right)= \frac{1}{4}-\frac{1}{2}(|a|^3 |b|-|a| |b|^3)\cos(\Delta\phi) \\
\nonumber
&\frac{1}{4} \left(1-q^1+q^0\right)= \frac{1}{4}-\frac{1}{2}(|a|^3 |b|-|a| |b|^3)\cos(\Delta\phi)\\
\nonumber
 &\frac{1}{4} \left( 1-q^1+q^0\right)= \frac{1}{4}+\frac{1}{2}(|a|^3 |b|-|a| |b|^3)\cos(\Delta\phi),
\end{align}
which clearly do not have a solution unless $(|a|^3 |b|-|a| |b|^3)\cos(\Delta\phi)=0$. Analogous to the case of the simple Wigner's friend experiment we can obtain solutions by increasing the number of variables. This will now introduce a dependence on Bob's result and lead to the modified set of equations
\begin{align}
\label{A2joint00}
&(1-q^{00})|\alpha|^2 |\nu|^2+q^{10}|\beta|^2|\mu|^2= \\
	&\qquad \qquad |\alpha|^2 |\nu|^2 (|a|^4+|b|^4)+ 2 |\beta|^2 |\mu|^2|a|^2|b|^2 +2\xi\nonumber \\
\label{A2joint01}
&(1-q^{01})|\alpha|^2 |\mu|^2+q^{11}|\beta|^2|\nu|^2= \\
 	&\qquad \qquad |\alpha|^2 |\mu|^2 (|a|^4+|b|^4)+ 2 |\beta|^2 |\nu|^2|a|^2|b|^2 - 2\xi \nonumber \\
 \label{A2joint10}
&(1-q^{10})|\beta|^2|\mu|^2 +q^{00}|\alpha|^2 |\nu|^2 =\\
	&\qquad \qquad |\beta|^2 |\mu|^2 (|a|^4+|b|^4)+2 |\alpha|^2 |\nu|^2|a|^2|b|^2 - 2\xi \nonumber \\
\label{A2joint11}
 &(1-q^{11})|\beta|^2|\nu|^2+q^{01}|\alpha|^2 |\mu|^2 =\\
 	& \qquad \qquad |\beta|^2 |\nu|^2 (|a|^4+|b|^4)+ 2 |\alpha|^2 |\mu|^2|a|^2|b|^2 +2\xi,\nonumber   
\end{align}
which can be solved for flipping probabilities $q^{nm}$. Again, the solutions are now no longer unique and allow for introducing additional requirements, for example that the dependence on Bob's result is as small as possible, i.e. $q^{n0}=q^{n1}+\varepsilon$ with minimal $|\varepsilon|$. Like in the simple case this requirement ensures that if there are solutions to Eqs.~\eqref{Ajoint00}-~\eqref{Ajoint11}, we recover them when solving Eqs.~\eqref{A2joint00}-~\eqref{A2joint11}. 

In general, all these solutions depend on Bob's measurement setting and might, therefore, lead to signaling. Note, however, that Wigner and his friend do not necessarily have access to Bob's result and may then at most access effective flipping probabilities
\begin{align}
 &\bar{q}^0= p(B_2=0)q^{00}+ p(B_2=1)q^{01}\\
 &\bar{q}^1= p(B_2=0)q^{10}+ p(B_2=1)q^{11}.
\end{align}
As long as the these effective probabilities are the same for all measurement settings of Bob, while solving Eqs.~\eqref{A2joint00}-~\eqref{A2joint11} or their simpler counterparts Eqs.~\eqref{Ajoint00}-~\eqref{Ajoint11}, the changes in the friend's memory register cannot be used to signal from Bob to Wigner and his friend. As we will show below, this cannot always be the case. 

First consider the scenario where the source emits the maximally entangled state (i.e. $\alpha=\beta=1/\sqrt{2}$ in Eq.~\eqref{initial_state}) and Bob measures in the computational basis, i.e. $\mu=1$, $\nu=0$. Eqs.~\eqref{A2joint00}-~\eqref{A2joint11} now simply read
\begin{align}
\nonumber
&\frac{1}{2}q^{10}=|a|^2|b|^2 \\
\nonumber
&\frac{1}{2}(1-q^{10})= \frac{1}{2} (|a|^4+|b|^4)\\
\nonumber
&\frac{1}{2}(1-q^{01})= \frac{1}{2} (|a|^4+|b|^4)\\
\nonumber
&\frac{1}{2}q^{01}=|a|^2|b|^2
\end{align}
and are solved by $q^{01}= \bar{q}^0=2|a|^2|b|^2=q^{10}=\bar{q}^1$. Hence, all other measurement setting by Bob must give the same effective flipping probabilities to avoid signalling. Now, consider the same scenario but Bob measuring in the basis given by $\mu= 1/\sqrt{3},\nu= \sqrt{2/3}$. In this case Eqs.~\eqref{A2joint00}-~\eqref{A2joint11} become 
\begin{align*}
\nonumber
&\frac{1}{6}q^{10}-\frac{1}{3}q^{00}\\
&=\frac{1}{3} \left(\sqrt{2}(|a|^3|b|-|a| |b|^3)\cos(\Delta\phi) - |a|^2|b|^2\right) \\
\nonumber
&\frac{1}{3}q^{11}-\frac{1}{6}q^{01}\\
&=\frac{1}{3} \left( |a|^2|b|^2-\sqrt{2}(|a|^3|b|-|a| |b|^3)\cos(\Delta\phi) \right) \\
\nonumber
&\frac{1}{3}q^{00}-\frac{1}{6}q^{10}\\
&=\frac{1}{3} \left( |a|^2|b|^2-\sqrt{2}(|a|^3|b|-|a| |b|^3)\cos(\Delta\phi) \right)\\
\nonumber
&\frac{1}{6}q^{01}-\frac{1}{3}q^{11}\\
&=\frac{1}{3} \left(\sqrt{2}(|a|^3|b|-|a| |b|^3)\cos(\Delta\phi)- |a|^2|b|^2\right),
\end{align*}
with $\Delta\phi=\phi_b-\phi_a$. These equations only determine the flipping probabilities of the friend's memory up to
\begin{align}
\label{cond1}
&2q^{00}-q^{10}=2 \left( |a|^2|b|^2-\sqrt{2}(|a|^3|b|+|a| |b|^3)\cos(\Delta\phi) \right) \\
\label{cond2}
&2q^{11}-q^{01}=2 \left( |a|^2|b|^2-\sqrt{2}(|a|^3|b|+|a| |b|^3)\cos(\Delta\phi) \right). 
\end{align}
Now assume that there exist solutions $q^{nm}$, such that the flipping of the friend's memory would also  be non signaling for any $a,b$. This means that in addition to Eqs.~\eqref{cond1} -~\eqref{cond2} we further require that
\begin{align}
\label{NScond1}
&q^{00}+q^{01}=4|a|^2|b|^2 \\
\label{NScond2}
&q^{11}+q^{10}=4|a|^2|b|^2. 
\end{align}
Plugging Eqs.~\eqref{NScond1} -~\eqref{NScond2} into Eqs.~\eqref{cond1} -~\eqref{cond2} gives
\begin{align}
\label{Condi}
q^{00}=2|a|^2|b|^2-\frac{2\sqrt{2}}{3}(|a|^3|b|-|a| |b|^3)\cos(\Delta\phi)=q^{11}.
\end{align}
As shown in Fig.~\ref{signaling} there exist measurements by Wigner (i.e. values of $a,b$) such that these solutions are negative and, therefore, not proababiliteis. This, in turn, means that for some settings solutions to Eqs.~\eqref{cond1} and~\eqref{cond2} that are probabilities cannot simultaneously fulfill Eqs.~\eqref{NScond1} and~\eqref{NScond2}. Hence, for these settings the effective flipping probabilities have to be signaling. \\

\end{appendix}

\end{document}